\documentclass [prl,reprint,notitlepage,superscriptaddress,floatfix] {revtex4-1}
\usepackage{amsmath}
\usepackage{graphicx}
\usepackage{lmodern}
\usepackage{amsmath}
\usepackage{hyperref}

\usepackage{color}
\usepackage{amssymb}
\usepackage{bm}
\usepackage{ulem}
\newcommand{\beq} {\begin{equation}}
\newcommand{\eeq} {\end{equation}}
\newcommand{\bea} {\begin{eqnarray}}
\newcommand{\eea} {\end{eqnarray}}
\newcommand{\be} {\begin{equation}}
\newcommand{\ee} {\end{equation}}

\DeclareMathOperator{\Ree}{Re}
\DeclareMathOperator{\Imm}{Im}

\begin{document}

\title {Superconductivity above a quantum critical point in a metal -- gap closing vs gap filling, Fermi arcs, and pseudogap behavior.}
\author{Artem Abanov}
\affiliation{Department of Physics, Texas A\&M University, College Station,  USA}
\author{Yi-Ming Wu}
\affiliation{School of Physics and Astronomy and William I. Fine Theoretical Physics Institute, University of Minnesota, Minneapolis, MN 55455, USA}
\author{Yuxuan Wang}
\affiliation{Department of Physics, University of Florida, Gainesville, FL 32611,  USA}
\author{Andrey V. Chubukov}
\affiliation{School of Physics and Astronomy and William I. Fine Theoretical Physics Institute, University of Minnesota, Minneapolis, MN 55455, USA}
\date{\today}

\begin{abstract}
We consider  a quantum-critical metal with interaction mediated by  fluctuations of a critical order parameter. This interaction gives rise to two competing tendencies -- pairing and  non-Fermi liquid behavior.    Due to competition, the  pairing develops below a finite
$T_p $,  however its prominent feedback  on the fermionic self-energy develops only at a lower $ T_{cross}$.  At $T<T_{cross}$ the system behavior  is similar to that of a BCS supercoductor -- the density of states (DOS) and the spectral function (SF) have  sharp gaps which  {\it close} as $T$ increases. At higher $T_{cross}<T<T_{p}$ the DOS has a dip, which  {\it fills in} with increasing $T$. The SF in this region  shows either the  same behavior as the DOS, or has a peak at $\omega =0$ (the Fermi arc), depending on the position on the Fermi surface. We argue that  phase fluctuations are strong in this $T$ range, and  the actual $T_c \sim T_{cross}$, while $T_p$ marks the onset of pseugogap behavior. We
 compare our theory with the behavior of optimally doped  cuprates.
\end{abstract}
\maketitle

{\it Introduction.}~ Pairing near a quantum-critical point (QCP) in a metal and its intriguing interplay with the concurring non-Fermi liquid (NFL) physics
  continue
  to attract strong attention of the physics community~\cite{scal,*scal2,acf,*acs,book1,review3,review4,mack,varma,matsuda}.  An incoherence associated with the NFL form of the self-energy, acts against pairing, while the latter reduces fermionic incoherence by  gapping out low-energy states.  The competition between these two opposite tendencies has been analyzed analytically, using field-theoretical methods for effective low-energy models~~\cite{combescot,*Bergmann,*Bergmann2,*ad,*Marsiglio_88,*Marsiglio_91,*Karakozov_91,nick_b,acf,*acs,son,sslee,
subir2,*moon_2,max2,wang_2,*wang23,max_last,raghu_15,mack,scal,*scal2,efetov,steve_sam,Wang2016,tsvelik,vojta,khvesh,Kotliar2018,*we_last_D}, and numerically, by, e.g., FRG,  QMC and DMFT techniques~\cite{metzner,berg,*berg_2,*berg_3,kotliar,*kotliar2,review3,*tremblay_2,georges,*georges2}.
Earlier  studies have found~\cite{combescot,*Bergmann,*Bergmann2,*ad,*Marsiglio_88,*Marsiglio_91,*Karakozov_91,acf,*acs,max_last,raghu_15,Wang2016,Kotliar2018,*we_last_D} that the onset temperature for the pairing, $T_p$, is finite, i.e., a QCP is surrounded by a superconducting done.

\begin{figure}
      \begin{center}
        \includegraphics[width=8.5cm]{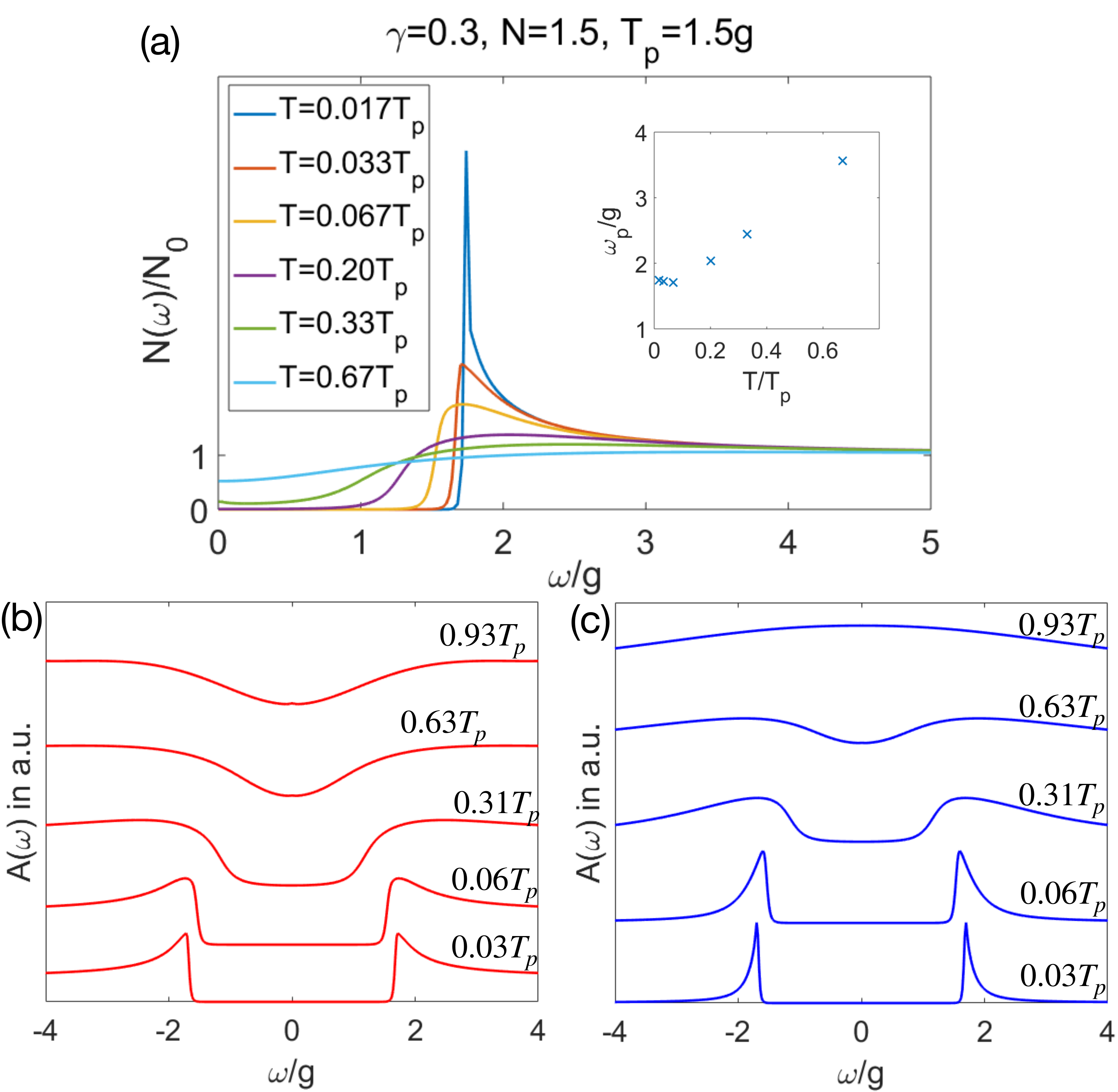}
        \caption{The DOS (a) and the SF (b and c)  at various $T < T_p$.  For the SF, (b) and (c) are for strong and weak
        thermal contribution from static interaction, which for the cuprates we associate with antinodal and near-nodal regions on the FS.
         At low $T < T_{cross}$, both DOS and SF have the peaks at
        $|\omega| \approx \Delta (T)$, and the peak frequency decreases as $T$ increases, i.e. the gap  "closes''.  At $T_{cross} <T < T_{p}$  the DOS and the SF in panel (b) show a dip at small $\omega$ and a hump, whose position increases with increasing $T$.  At $T$ approaches $T_c$, the DOS and the SF  flatten up, i.e., the gap "fills in''. The SF in panel (c) shows a single peak at $\omega =0$ instead of a dip (a Fermi arc).}
        \label{fig:summary}
      \end{center}
    \end{figure}

  The issue which we discuss here  is the feedback on the fermions from the pairing in the quantum critical (QC) regime, specifically the behavior of the DOS $N(\omega)$ and the spectral function $A_{{\bf k}_F}(\omega)$. We argue that there are two distinct regimes below $T_p$, which differ by the strength of the feedback from the pairing on the electrons.   At low  $T < T_{cross} < T_p$ (regime I) the feedback is strong, and both $N(\omega)$ and $A_{{\bf k}_F}(\omega)$ have sharp quasiparticle peaks at $\omega = \Delta (T)$.  At higher $ T_{cross} <T < T_{p}$ (regime II) the feedback is weak, and $N(\omega)$ has a dip at $\omega =0$ and a hump at a frequency which scales with $T$ rather than $\Delta (T)$
   and remains finite at $T=T_p$. The behavior of the  spectral function (SF)  $A_{{\bf k}_F} (\omega)$  in this region depends on the strength of thermal contribution to the self-energy and  varies  along the Fermi surface (FS).  For ${\bf k}_F$  points where it is strong enough,  $A_{{\bf k}_F} (\omega)$ displays the same behavior as $N(\omega)$,  for other ${\bf k}_F$ points, $A_{{\bf k}_F} (\omega)$  shows a peak at $\omega =0$ instead of a dip (the Fermi arc).  We summarize the results in Fig. \ref{fig:summary}
A very similar  evolution of both DOS and SF has been observed in the cuprates around optimal doping~\cite{DOS,dessau,kaminski,kanigel,*kanigel2,*kanigel3,norman_review,shen,*shen2,*shen3,*shen4,hoffman} and dubbed as transformation from ``gap closing'' at low $T$ to ``gap filling in'' at higher $T$. The  behavior similar to Figs. 1a and 1b  has been detected in tunneling measurements of $N(\omega)$ and ARPES measurements of $A_{{\bf k}_F} (\omega)$ at ${\bf k}$ near $(\pi,0)$ and related points in the Brillouin zone (BZ).  The behavior similar to Fig. 1c
  has been found in ARPES measurements closer to zone diagonals. Our reasoning for the difference between Figs 1b and c  is valid if the pairing  boson predominantly couples fermions near $(\pi,0)$ and related points, like a $(\pi,\pi)$  spin fluctuation does~\cite{scal,*scal2,acf,*acs}

We further analyze the superfluid phase stiffness $\rho_s (T)$ and argue that in regime II, $\rho_s (T) < T$, that is,
  phase fluctuations are strong and likely destroy long range superconducting order~\cite{emery_kiv,*mohit_1}. Then the actual $T_c \sim T_{cross}$, while between $T_{cross}$ and $T_p$  the system displays  pseudogap behavior. This also agrees with the experiments, which found  that the crossover  from gap closing to gap filling  occurs at around $T_c$ (Ref. \cite{dessau}), while the dip in $N(\omega)$ at $\omega =0$ disappears at a higher $T$.  It is tempting to associate the onset of the pseudogap behavior in the cuprates at optimal doping  with our $T_p$.  We caution that we do not associate the whole pseudogap region in the cuprates with the pairing state  with no phase coherence and with  weak feedback on the fermionic self-energy.  There are other features, like apparent FS reconstruction~\cite{taillefer,subir_last}, charge and nematic orders~\cite{Fujita,*cdw1,*cdw2,*cdw3,*cdw4,*cdw5,*cdw6,*cdw7,italy,*sachdev2013,*wang_ch}, and time-reversal symmetry breaking~\cite{aharon,greven}, which we do not address in this study. Still, we believe that this work presents microscopic understanding of ``weak'' pseudogap behavior --- the crossover from gap closing to gap filling with increasing $T$ in the DOS and the SF in the anti-nodal regions of the FS, and the development of Fermi arcs in the SF in near-nodal regions.

  We also note that the issue we consider here is different from peak-dip-hump in the SF and related phenomena in the DOS, optical conductivity, and other measurements~\cite{fink}.
   That phenomenon has been associated with the emergence of quasiparticle scattering in a superconductor at frequencies somewhat below $3\Delta$, well separated from the peak at $\omega =\Delta$. The peak-dip-hump phenomenon was explained in terms of coupling to a propagating boson, either a phonon~\cite{devereaux}, or a spin fluctuation -- the latter becomes propagating below $T_p$ due to the development of a resonance mode below $2\Delta$
    (Ref. \cite{eschrig,*abanov_ch}).  The phenomenon we discuss here is the destruction of the peak at $\Delta$ due to the existence, in regime II, of strong quasiparticle scattering  down to the lowest energies. This phenomenon has been described phenomenologically in the past~\cite{imp_general,*imp_general2,*imp_general3,imp_cuprates,*imp_cuprates2,*imp_cuprates3,*imp_cuprates4,*imp_cuprates5,elihu}, by introducing  frequency independent fermionic damping $\gamma (T)$ and allowing it to be comparable to the gap $\Delta (T)$.  Our work presents the microscopic theory of the existence of  $\text{Im} \Sigma (\omega \to 0, T)$ at  $T > T_{cross}$, despite the fact that the pairing gap is non-zero.

 {\it The model.}~We consider the model of itinerant fermions minimally coupled to fluctuations of the order parameter field, which condenses at a QCP. Within Eliashberg-type approximation, which we adopt, the effective 4-fermion interaction is proportional to the   susceptibility of an order parameter integrated along the FS, $\chi (\Omega_m)$.  At a QCP,  $\chi (\Omega_m) =(g/|\Omega_m|)^{\gamma}$ is a singular function of frequency
  (the exponent $\gamma$ is small near 3D,  and in 2D equals to $1/3$ at a nematic QCP  and to $1/2$ at QCP towards
   a
   density-wave order~\cite{nick_b,max_last,acf,*acs,raghu_15,Wang2016}).  The regimes I and II exist for all  $\gamma <1$, and below we do not specify the value of the exponent.
   The singular $\chi (\Omega_m)$  gives rise to an attraction in at least one pairing channel and  also gives rise to NFL behavior in the normal state, setting the competition between the pairing and the NFL behavior.  We consider spin-singlet pairing and solve the set of non-linear equations for the pairing vertex  and fermionic self-energy on the Matsubara axis, and then convert the results to  real frequencies~\cite{combescot,*Bergmann,*Bergmann2,*ad,*Marsiglio_88,*Marsiglio_91,*Karakozov_91} and find the DOS and the SF. We present the details of the calculations in~\cite{last} and here show the results.

Along the Matsubara axis, the coupled equations for the pairing vertex $\Phi (\omega_m)$ and fermionic self-energy $\Sigma (\omega_m)$ are~\cite{Wang2016} (${\tilde \Sigma} (\omega_m) = \omega_m + \Sigma (\omega_m)$)
\begin{eqnarray}\label{eq:gapeq}
&&    \Phi (\omega_m) =
    \pi T  \sum_{m'} \frac{\Phi (\omega_{m'})\chi (\omega_{m}-\omega_{m'})}{\sqrt{{\tilde \Sigma}^2 (\omega_{m'}) +\Phi^2 (\omega_{m'})}}  \\
&&     {\tilde \Sigma} (\omega_m) = \omega_m
   +  g^\gamma
    \pi T \sum_{m'}  \frac{{\tilde \Sigma}(\omega_m)\chi (\omega_{m}-\omega_{m'})}{\sqrt{{\tilde \Sigma}^2 (\omega_{m'})  +\Phi^2 (\omega_{m'})}}, \nonumber
\end{eqnarray}
In principle, one should also include the equation for bosonic self-energy, which describes the feedback from $\Phi (\omega_m)$ on $\chi (\Omega_m)$ (Refs. \cite{eschrig,*abanov_ch}).
This feedback effectively makes the exponent $\gamma$ temperature dependent below $T_p$.  However, because regimes I and II exist for all $\gamma$, this will only affect the location of $T_{cross}$. Below we neglect this complication and treat the exponent $\gamma$ as temperature independent.

The thermal contributions to $\Phi (\omega_m)$ and  ${\tilde \Sigma} (\omega_m)$ come  from  $m' = m$  terms in the sums. These contributions are essential for the SF, but can be  excluded from the Eliashberg set by analogy with non-magnetic impurities~\cite{agd,msv,acn},  by re-expressing
$\Phi (\omega_m) = \Phi^{\ast} (\omega_m) \left(1+ Q^{\ast} (\omega_m)\right), ~{\tilde \Sigma} (\omega_m) = {\tilde \Sigma}^{\ast} (\omega_m) \left(1+ Q^{\ast} (\omega_m)\right)$, where $Q^{\ast} (\omega_m) =  \pi T \chi (0)/\sqrt{({\tilde \Sigma}^{\ast})^2 (\omega_{m}) +(\Phi^{\ast} (\omega_{m}))^2}$.
 The equations for $\Phi^{\ast}$ and ${\tilde \Sigma}^{\ast}$ are the same as in (\ref{eq:gapeq}) but without $m=m'$ term in the sum. We solve these two equations and then obtain $\Phi^{\ast} (\omega)$ and ${\tilde \Sigma}^{\ast} (\omega)$ using spectral decomposition method and analytical continuation. In the normal state (at $\Phi^\ast =0$) the self-energy has a NFL form $\Sigma^\ast (\omega_m) \propto \omega^{1-\gamma}_m$.

  The onset temperature for the pairing $T_p = T_p (\gamma)$ has been obtained in~\cite{raghu_15,Wang2016}.  It is finite, of order $g$, and scales as $\gamma^{-1/\gamma}$ at small $\gamma$.
The pairing gap $\Delta (\omega)$ is defined as $\Delta (\omega) = \Phi (\omega) \omega/{\tilde \Sigma} (\omega)$ and is equally expressed as
$\Delta (\omega) = \Phi^{\ast} (\omega) \omega_m/{\tilde \Sigma}^{\ast} (\omega)$.  The DOS $N(\omega)$ is expressed only via $\Delta (\omega)$: $N(\omega) = N_0 \Imm \omega/\sqrt{\Delta^2 (\omega) - \omega^2}$ and therefore has no contribution from thermal fluctuations. The SF $A(\omega)$ for a fermion on the FS does  depend on the thermal contribution: $A(\omega >0) =
 (-1/\pi) \Imm L (\omega) \omega/\sqrt{\Delta^2(\omega)-\omega^2}$, where
 $L(\omega) =  1/(\pi T \chi (0) - ({\tilde \Sigma} (\omega)/\omega)\sqrt{\Delta^2 (\omega)-\omega^2})$.

\begin{figure}
	\begin{center}
  \includegraphics[width=8.8cm]{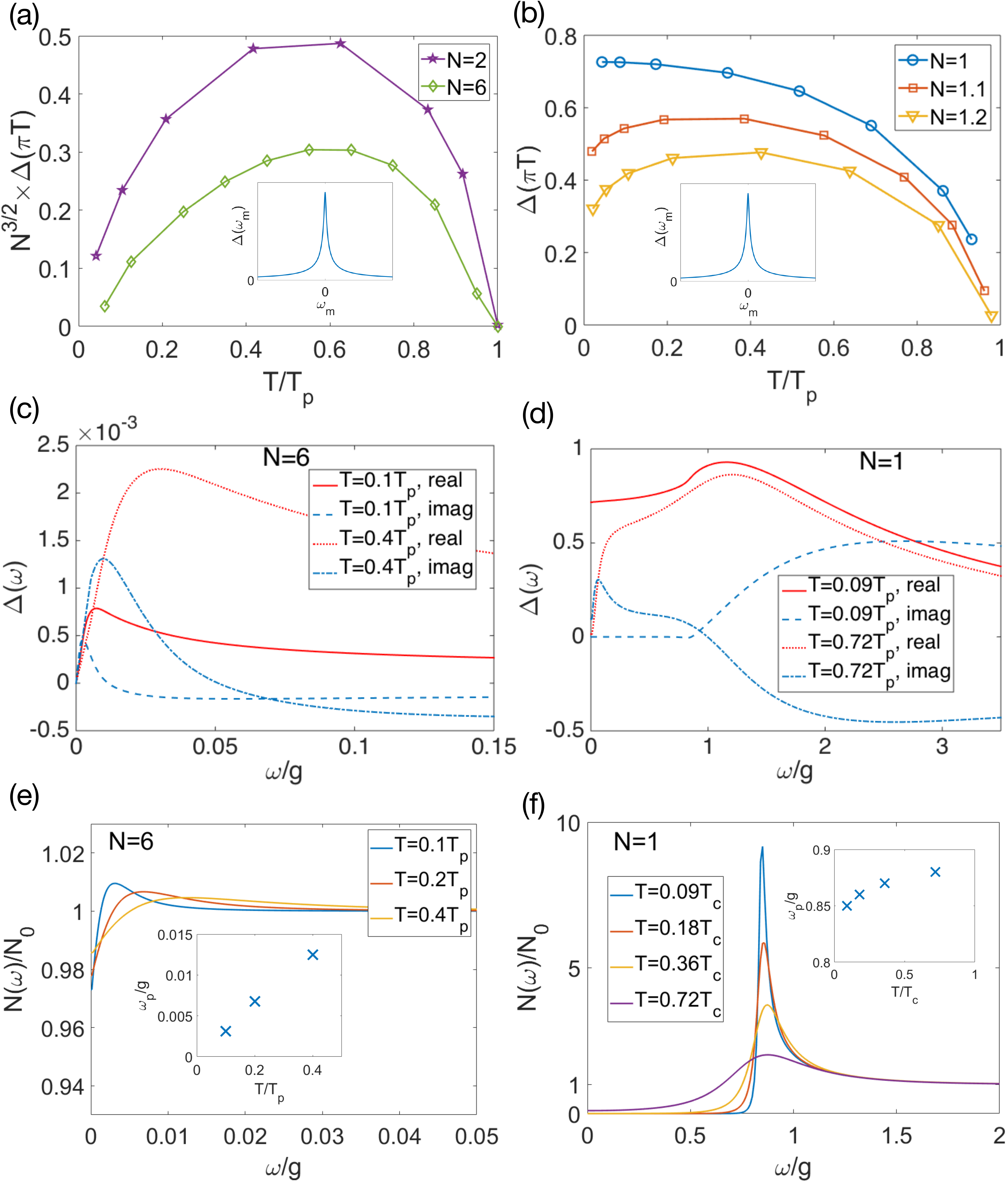}
    	\caption{
      Numerical results for $\gamma=0.9$ ($N_{cr} =1.34$). 
      (a) The gap at the first Matsubara frequency $\Delta (\pi T)$ (in units of g)  as a function of temperature for several $N>N_{cr}$. The gap function is non-monotonic and vanishes at $T=T_p$ and $T=0$.
 Inset: $\Delta (\omega_m)$ as a function of $\omega_m$ at a given $T$. (b)  $\Delta (\pi T) $ as a function of temperature for several $N<N_{cr}$.
 (c) and (d):$\Ree \Delta(\omega)$  and $\Imm \Delta (\omega)$  as functions of real frequency $\omega$ for $N=6$ and $N=1$ ($N > N_{cr}$ and $N < N_{cr}$).
 (e) and (f): The DOS $N(\omega)$ for various T for $N=6$ and $N=1$. Insets: the position of the maximum (hump)$\omega_p$ vs $T$. }\label{fig:F}
	\end{center}
\end{figure}

\begin{figure}
	\begin{center}
		\includegraphics[width=8cm]{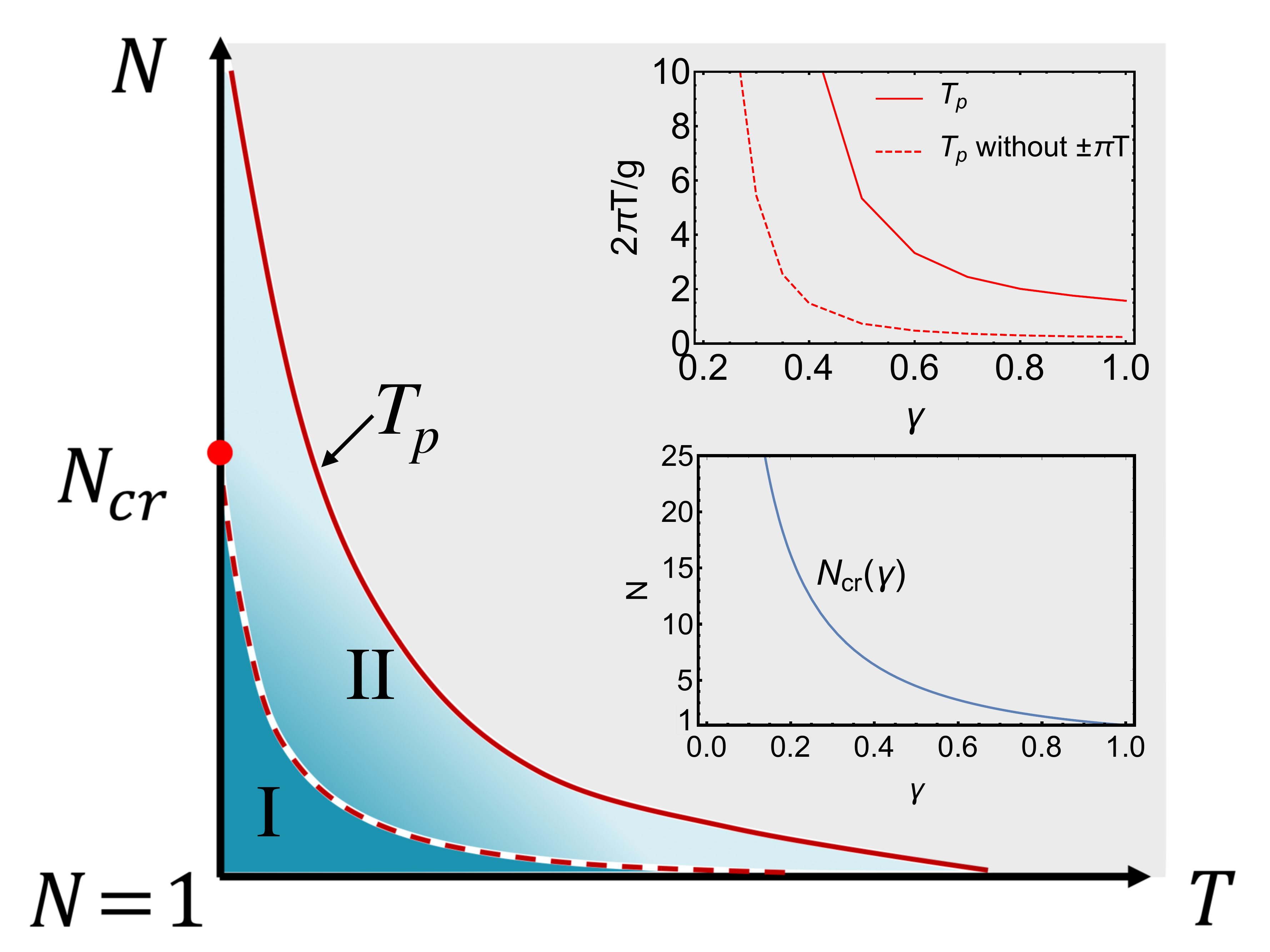}
		\caption{The phase  diagram of our QC model, extended to $N >1$ (see text), for some $\gamma<1$.
   The solid line is the onset temperature for the pairing, $T_p (N)$. The dashed line  marks the crossover between BCS-type  behavior at $T < T_{cross}$ (region I) and the novel behavior at a higher $T$ (region II), in which  fermionic self-energy remains approximately the same as in the normal state.  Phase fluctuations
      likely destroy long-range phase coherence in phase II, in which case the actual $T_c \sim T_{cross}$ and region II is the pseudogap phase.  Insets: the onset temperature for the pairing with and without fermions with first Matsubara frequency $\omega_m = \pm \pi T$ for $N=1$, and $N_{cr} (\gamma)$. }\label{fig:newT}
	\end{center}
\end{figure}

{\it The two regimes below $T_p$.}~
The existence of the two different regimes below $T_p$ can be understood by analyzing the gap equation along the Matsubara axis. We argue
  that the existence of the regimes I and II is associated with the special role of fermions with Matsubara frequencies $\omega_m = \pm \pi T$. Namely,  fermionic self-energy $\Sigma^{\ast} (\omega_m)$,
which acts against the pairing,  is strong and singular at a QCP for a generic $\omega_m$ but  vanishes at $\omega_m = \pm \pi T$, i.e. fermions with these two frequencies can be treated  as free particles for the purposes of the pairing~\cite{Wang2016,maslov}.   Meanwhile, the pairing interaction between fermions with $\omega_m = \pi T$ and $\omega_m = - \pi T$, $\chi (2\pi T) = (g/(2\pi T))^\gamma$ is strong. As the consequence, fermions with $\pm \pi T$   form a bound pair at $T_p$, and in some range below $T_p$ (in region II) act as the source for the pairing gap for fermions with other Matsubara frequencies.   At a smaller $T < T_{cross}$ (regime I)  fermions with other Matsubara frequencies become capable to pair on their own, without an input from fermions with $\omega_m = \pm \pi T$. We verify this by solving the gap equation with and without fermions with $\omega_m = \pm \pi T$. We find  (see the inset to Fig. \ref{fig:newT}) that in the first case  the critical temperature is $T_p$, and in the second it is a smaller $T_{cross}$.
  The special role of fermions with $\omega_m = \pm \pi T$  becomes more transparent if we modify the original model and reduce the
 the interaction in the pairing channel by $1/N$ compared to that  for the self-energy. This can be achieved by extending the model to $SU(N)$ global symmetry~\cite{raghu_15}.  At $N$ larger than some $N_{cr} (\gamma) \equiv  (1-\gamma )\Gamma (\gamma /2)\left(\frac{\Gamma (\gamma /2)}{2\Gamma (\gamma )}+\frac{\Gamma (1-\gamma )}{\Gamma (1-\gamma /2)} \right)>1$, $T_p$ is finite only due to fermions with $\omega_m = \pm \pi T$, i.e.,  the regime II extends down to $T=0$ (the line $T_{cross} (N)$ terminates at $N = N_{cr}$), see Fig. \ref{fig:newT}).  This extension
 does not change the physics as the regime II exists already for the original $N=1$, but it allows us to analyze the system behavior analytically, in $1/N$ expansion.
   We find that $\Phi^\ast (\omega_m)$ and $\Sigma^\ast (\omega_m)$ are proportional to $T^{1-\gamma}$, i.e., the self-energy retains its NFL form. The pairing gap $\Delta (\pi T) = \pi T (2/N)^{1/2} \left(1- (T/T_p)^\gamma\right)^{1/2}$ is small in $1/N$, and for other $\omega_m$, $\Delta (\omega_m) \propto 1/N^{3/2}$ is even smaller.  For all $\omega_m$, the  gap $\Delta (\omega_m)$  emerges at $T_p$ and vanishes at $T=0$,
     consistent with the fact that it is induced by fermions with $\omega_m = \pm \pi T$ and wouldn't exist without them. We show $\Delta (\omega_m)$ in Fig. \ref{fig:F}(a),(e).

The large-$N$ expressions for $\Phi^{\ast} (\omega_m)$ and  $\Sigma^{\ast} (\omega_m)$ can be analytically
converted to real frequencies. We obtain
 \begin{eqnarray}
&&  \Phi^* (\omega) =  \left(\frac{2}{N}\right)^{3/2}\!\!\!\!\!\!\! \pi T  \left(\frac{g}{\pi T}\right)^\gamma  \left[1 - \left(\frac{T}{T_p}\right)^\gamma\right]^{1/2}\!\!\!\!\!\!\!
  F_\Phi \left(\frac{\omega}{\pi T}\right) \nonumber \\
&& \tilde{\Sigma}^{\ast}(\omega)  = \pi T  \left(\frac{g}{\pi T}\right)^\gamma  F_\Sigma \left(\frac{\omega}{\pi T}\right).
 \label{s_18}
\end{eqnarray}
where $F_\Phi$ and $F_\Sigma$ are two scaling functions which only depend on $\omega/(\pi T)$\cite{last}. We see that the frequency dependence of both $\Phi^* (\omega)$ and ${\tilde \Sigma}^* (\omega)$  (an{}d hence of $\Delta (\omega)$ and the DOS and the SF in the anti-nodal region) is set by $T$ rather than by the gap, hence the position of the maximum  scales with $T$ and  remains
finite at $T_p$. As the consequence, the  gap fills in as $T \to T_p$, but does not close (Fig. \ref{fig:F}(b),(c)). At the smallest frequencies
$\Sigma^\ast  (\omega) \approx i \Imm \Sigma^\ast (0)$, and hence   $\Delta (\omega) \propto i \omega$, like for gapless superconductivity. Then
  the DOS $N(\omega)$ is then non-zero down to the lowest frequencies  (Fig. \ref{fig:F}(c)).

At smaller $N < N_{cr}$, including physical $N=1$, below $T_{cross}$ fermions with $|\omega_m| \neq \pi T$ can pair on their own, without a push from fermions with $|\omega_m| = \pi T$.   This can be checked by solving the Eliashberg equations without fermions with the first Matsubata frequencies (see the insert to Fig. \ref{fig:newT}).  Once other fermions get paired below $T_{cross}$,  the system recovers a conventional superconducting behavior, i.e., $\Delta (\pi T)$ tends to a finite value $\Delta$ at $T=0$, and at low $T$, the gap $\Delta (\omega)$ along the real axis is purely real at $\omega \leq \Delta$ (see Fig. \ref{fig:F}(e),(f)). This is region I in our notations.  Because the scattering at small $\omega$ is destroyed by a feedback from the pairing,  the DOS and the SF have sharp peaks at $\omega = \Delta$ (see the lowest $T$ data in Fig. \ref{fig:summary}).  As temperature increases but remains smaller than $T_{cross}$, the position of the maximum follows $\Delta (T)$ and decreases, i.e., the gap start closing. However, once $T$ exceeds $T_{cross}$, the system crosses over to the region II, where the pairing would not be possible without fermions with $\omega_m = \pm \pi T$, and the position of the hump in the DOS and the SF is set by $T$ rather than $\Delta (T)$.
 In this region, the gap progressively fills in as $T$ approaches $T_p$, but does not close. We show the results in  Fig. \ref{fig:summary}).

 {\it Superfluid stiffness.}~ Our next goal is to verify whether in region II  superconducting order may be destroyed by phase fluctuations. Our  earlier results do not rely on phase coherence  and are applicable even in the absence of long-range phase coherence.
  Moreover, the feedback effect from the pairing on fermionic self-energy is weak in region II already  in the absence of  phase fluctuations, and phase decoherence can only reduce the feedback even further. Still, for comparison with the cuprates it is important to understand whether the region II corresponds to  superconducting or pseudogap phase.

  To compute the stiffness we follow Ref. \cite{cee}, set $\Delta (\omega_m,r) = \Delta (\omega_m) e^{i {\bf \nabla} \phi {\bf r}}$, i.e.,
    $\Delta (\omega_m, q) = \Delta (\omega_m) \delta ({\bf q}- {\bf \nabla} \phi)$, and identify $\rho_s (T)$ with the prefactor for the $q^2$ term in the condensation energy. For a BCS superconductor  $\rho_s (T<T_c) \approx E_F/(4\pi)$.  Because $E_F$ is assumed to be much larger than $T_c$, phase fluctuations are weak.  In our case, we found at large $N$, when region II extends to $T=0$,
    \beq
    \rho_s (T) \approx \frac{T}{N} \left(1- \left(\frac{T}{T_p}\right)^\gamma\right)  \frac{E_F}{\pi T \chi (0)} \left(1 + O\left(\frac{1}{N}\right)\right),
    \eeq
    where, we remind, $\chi (0)$ is a static susceptibility of a critical bosonic field. It diverges at a QCP, so formally $\rho_s  (T)\to 0$.  However, Eliashberg theory is only valid when $E_F \geq \pi T \chi (0)$ because  the integration over fermionic dispersion  only holds up to $E_F$. This restricts  $\chi (0)$  to $\pi T \chi_0 \leq E_F$ and $\rho_s$ to $\rho_s (T) \geq (T/N) \left(1- \left(T/T_p\right)^\gamma\right)$. Still, at large $N$, $\rho_s (T)\ll T$. In this situation, phase fluctuations are strong and should destroy long range phase coherence
    ~\cite{emery_kiv,*mohit_1,benfatto}.
     Then region II becomes the pseudogap phase. In region I the same calculation yields $\rho_s \geq \Delta (T=0)$, i.e., phase fluctuations are at most moderate and phase coherence survives.

   {\it Summary.}~ In this paper we analyzed the feedback on the fermions from the pairing in the QC regime, specifically the behavior of the DOS $N(\omega)$  and the SF $A_{{\bf k}_F} (\omega)$ on the FS.  We considered the model of 2D fermions with singular interaction mediated by the dynamical susceptibility of a critical boson $\chi (\Omega_m) = (g/|\Omega_m|)^\gamma$.  This interaction gives rise to pairing and also to NFL behavior, which competes with the pairing.  To separate between the two tendencies, we extended the model in such a way that the  pairing interaction gets smaller by $1/N$.

    Our results are summarized in Fig. \ref{fig:newT}.  We found two distinct regimes below the onset temperature for the pairing $T_p$. They differ in the strength of the feedback from superconductivity on the electrons.   At low  $T < T_{cross} < T_p$ (regime I) the feedback is strong, and both $N(\omega)$ and $A_{{\bf k}_F}(\omega)$ have sharp quasiparticle peaks at $\omega = \Delta (T)$.  At higher $ T_{cross} <T < T_{p}$ (regime II) the feedback is weak, and $N(\omega)$ has a dip at $\omega =0$ and a hump at a frequency, which scales with $T$ rather than $\Delta (T)$ and remains finite at $T=T_p$, i.e., near $T_p$ the gap fills in but does not close.   The SF either has the same structure as the DOS or a peak at $\omega=0$ (the Fermi arc), depending on the location on the FS.  We computed superfluid stiffness and estimated the strength of phase fluctuations.  We found that in region II phase fluctuations are parameterically strong, at least at large $N$, and destroy phase coherence. Then the actual $T_c \sim T_{cross}$, while in between $T_{cross}$ and $T_p$ the system is in phase-disordered pseudogap state.   A very similar behavior has been detected in tunneling and ARPES studies of the cuprates near optimal doping, and we  propose our theory as a microscopic explanation of the observed behavior. From theoretical perspective, we associate the existence of the region II with special role of fermions with Matsubara frequencies $\pm \pi T$, which appear in the pairing channel without self-energy and form a Cooper pair  even when the  pairing interaction is reduced by $1/N$. These fermions then induce pairing for fermions with other Matsubara frequencies.
      \acknowledgements
  We thank   D. Dessau, A. Millis, N. Prokofiev, S. Raghu, G. Torroba, and  A. Yazdani for useful discussions.   This work by Y. Wu and AVC was supported by the NSF DMR-1523036.

\bibliography{1stMref}

\end{document}